\documentclass[10pt,journal,compsoc]{IEEEtran}

\usepackage[T1]{fontenc}
\usepackage[english]{babel}
\usepackage{graphicx,caption}
\usepackage{subfigure}
\usepackage{amsthm}
\usepackage{float}
\usepackage{subfloat}
\usepackage{epsfig}
\usepackage{url}
\usepackage{multirow}
\usepackage{graphicx}
\usepackage{amsmath}
\usepackage{amssymb}
\usepackage{mathrsfs}
\usepackage{amsfonts}
\usepackage{paralist}
\usepackage{tabularx}
\usepackage{booktabs}
\usepackage{color,xcolor}
\usepackage{booktabs}

\usepackage{algorithm}
\usepackage{algpseudocode}

\newtheorem{thm}{Theorem}

\theoremstyle{remark}

\theoremstyle{definition}

\newfloat{problem}{htb}{loM}
\floatname{problem}{Problem}

\ifCLASSOPTIONcompsoc
  \usepackage[nocompress]{cite}
\else
  \usepackage{cite}
\fi
\ifCLASSINFOpdf
\else
\fi
\begin{document}
%
\graphicspath{{images/}}
\title{An Edge Computing-based Photo Crowdsourcing Framework for Real-time 3D Reconstruction}
%
%
%
%

\author{Shuai~Yu,~
        Xu~Chen,~
        Shuai~Wang,~
        Lingjun~Pu,~
        Di~Wu
\IEEEcompsocitemizethanks{\IEEEcompsocthanksitem Shuai Yu, Xu Chen and Di Wu are with the School of Data and Computer Science, Sun Yat-sen University, Guangzhou 510275, China (Email: yushuai, chenxu35 and wudi27@mail.sysu.edu.cn).
\IEEEcompsocthanksitem Shuai Wang is with the Department of Electrical and Electronic Engineering, Southern University of Science and Technology, Shenzhen, China (Email: wangs3@sustech.edu.cn).
\IEEEcompsocthanksitem Lingjun Pu is with the College of Computer Science, Nankai University, Tianjin, China (Email: pulingjun@nankai.edu.cn).
}


}

\IEEEtitleabstractindextext{%
\begin{abstract}
Image-based three-dimensional (3D) reconstruction utilizes a set of photos to build 3D model and can be widely used in many emerging applications such as augmented reality (AR) and disaster recovery.
Most of existing 3D reconstruction methods require a mobile user to walk around the target area and reconstruct objectives with a hand-held camera, which is inefficient and time-consuming.
To meet the requirements of delay intensive and resource hungry applications in 5G, we propose an edge computing-based photo crowdsourcing (EC-PCS) framework in this paper.
The main objective is to collect a set of representative photos from ubiquitous mobile and Internet of Things (IoT) devices at the network edge for real-time 3D model reconstruction, with network resource and monetary cost considerations.
Specifically, we first propose a photo pricing mechanism by jointly considering their freshness, resolution and data size.
Then, we design a novel photo selection scheme to dynamically select a set of photos with the required target coverage and the minimum monetary cost.
We prove the NP-hardness of such problem, and develop an efficient greedy-based approximation algorithm to obtain a near-optimal solution.
Moreover, an optimal network resource allocation scheme is presented, in order to minimize the maximum uploading delay of the selected photos to the edge server.
Finally, a 3D reconstruction algorithm and a 3D model caching scheme are performed by the edge server in real time.
Extensive experimental results based on real-world datasets demonstrate the superior performance of our EC-PCS system over the existing mechanisms.
\end{abstract}

}

\maketitle

\IEEEdisplaynontitleabstractindextext

%
\IEEEpeerreviewmaketitle

\IEEEraisesectionheading{\section{Introduction}\label{sec:introduction}}

%
%
%
%
\IEEEPARstart{I}{n} recent years, the advances in hardware technology have induced the emergence of mobile immersive applications (e.g., Augmented Reality (AR)~\cite{4161027} and 3D gaming~\cite{VR_game}).
Generally, these applications require accurate, large-scale, and dense 3D models of environments in (or close to) real-time, thus are delay intensive and resource hungry.
3D reconstruction~\cite{DBLP:journals/corr/abs-1906-06543, 1638022, 7335496} is a very well studied problem in computer vision.
Traditional image-based reconstruction techniques~\cite{7335496, PhotoCity} require a mobile user to capture multiple RGB images of the same object from multiple viewpoints.
Obviously, this may not be practical or time-consuming in large-scale outdoor environments.

One key challenge for image-based 3D reconstruction is to obtain a set of representative photos that best cover the target area.
As a novel sensing paradigm, photo crowdsourcing~\cite{8056963} which leverages the power of ordinary mobile users for large-scale sensing, has become popular in recent years.
With photo crowdsourcing, photos taken by participants can be collected and processed in cloud server, and thus facilitate a variety of image-based mobile applications.
However, reconstructing 3D models through cloud-based photo crowdsourcing still faces the following challenges:
\textit{i) Large photo uploading overhead.}
Participants may have tons of real-time captured photos for a target area and most of the photos may contain duplicated information.
Thus, uploading all the photos to remote cloud server will be unnecessary and consume tremendous network resources;
\textit{ii) Long photo collection delay.}
Achieving omnidirectional viewpoints of target area through ordinary mobile users is time-consuming.
The cloud server may wait a very long time until the photos with all viewpoints of target area are achieved;
\textit{iii) Limited shooting angles of participants.}
Ordinary mobile users with hand-held mobile devices are limited by their geographic location.
Some special shooting angles are impossible to reach.
For example, participants on the ground can not photograph the roof of a building;
and \textit{iv) Risk of privacy leakage.}
For the participants, their photos usually contain sensitive personal information (e.g., location and face information).
Thus, participants may hesitate to upload their photos to remote cloud server.

To this end, multi-access edge computing (MEC)~\cite{ETSIMEC,8736011,wang2019machine}, has been proposed by the European Telecommunications Standards Institute (ETSI), which is viewed as an ideal solution to address the above challenges.
In the MEC architecture, distributed MEC servers are deployed close to end users that provide cloud-computing capabilities and IT services with ultra-low latency, high bandwidth and privacy protection.

In this paper, we propose a novel framework for photo crowdsourcing in the 5G MEC network environments, in order to fully exploit the massive mobile and IoT devices (e.g., mobile phones, surveillance cameras, UAVs) and MEC resources in proximity for real-time 3D reconstruction.
Specifically, our edge computing-based photo crowdsourcing (EC-PCS) framework mainly consists of: i) a photo/participant selection module, that can select a set of photos with minimum monetary cost by leveraging photo crowdsourcing; ii) a 3D reconstruction module, that can reconstruct required 3D models based on the selected photos in the MEC server side; and iii) a data caching module, that can dynamically decide if a reconstructed 3D reconstruction should be cached in the MEC server.
The major contributions of this paper are summarized as follows:

\begin{itemize}
\item To guarantee good performance of real-time 3D reconstruction, it is essential to select sufficient and high quality fresh photos from participants in proximity (i.e., ubiquitous 5G mobile and IoT devices in this paper).
To this end, we first advocate a monetary-based incentive scheme by pricing photos according to its data size, resolution, freshness and the wireless channel states of their associated participants.
\item A photo selection scheme is proposed to select a representative set of photos with minimum monetary cost to satisfy the target coverage requirement.
While finding the optimal photo set with the minimum monetary cost is proven to be NP-hard, we propose a greedy-based approximation algorithm to obtain a near-optimal solution, and further theoretically characterize its approximation ratio.
\item When a representative set of photos is selected, the next question is how to efficiently utilize the limited network resources for fast photo uploading.
We thus design an optimal resource allocation scheme in order to minimize the maximum photo uploading delay.
\item Last but not least, the effectiveness of our EC-PCS framework is evaluated through extensive experiments based on real-world datasets, which demonstrates the superior real-time performance of EC-PCS solution over existing schemes.
\end{itemize}

The rest of this paper is organized as follows.
Section~\ref{sec:related_work} introduces the related works most relevant to this paper.
In Section~\ref{sec:System_Overview}, we present an overview for our proposed edge computing-based photo crowdsourcing (EC-PCS) framework.
Section~\ref{sec:Participant_Selection_Module} gives the details of the photo/participant selection module for our EC-PCS framework.
Section~\ref{sec:3D_Reconstruction_Module} introduces the 3D reconstruction module and the model caching module.
Simulation and experimental results are presented in Section~\ref{sec:performance_evaluation}.
Finally, conclusions are drawn in Section~\ref{sec:Conclusion}.

\section{Related Works}{\label{sec:related_work}}

\subsection{Real-time 3D Reconstruction}
Image-based 3D reconstruction technology~\cite{DBLP:journals/corr/abs-1906-06543} provides an efficient and low-cost solution to generate 3D models only with a set of photos from multiple viewpoints.
Although such reconstruction technology is less accurate compared to some laser scanning based technologies (e.g., simultaneous localization and mapping (SLAM)~\cite{1638022}), it is easier to be popularized with much less effort and cost.

In the 5G era, many emerging applications (e.g., augmented reality (AR)~\cite{4161027}) that require 3D model of objective can benefit from immediate feedback and low cost.
Thus, constructing 3D model in (or close to) real-time is critical for 5G-based mobile applications.
For example, authors in~\cite{6751117} propose an on-device live 3D reconstruction pipeline to meet the requirements of real-time 3D reconstruction scenarios (e.g., museums).
Most of existing real-time 3D reconstruction works are trajectory-based, which means that a mobile user walks around the target area with a hand-held device and reconstructs the environment.
For example, authors in~\cite{7335496} present an interactive 3D reconstruction system that can quickly build 3D models of indoor and large-scale outdoor environments.
In the system, a mobile user walks around a building with a Google Tango Tablet and reconstructs the scene, thus allowing the user to directly add data where it is needed.
Authors in~\cite{PhotoCity} design an online game PhotoCity that trains its players to take photos from multiple specific viewpoints and in great density.
The objective of such game is to create 3D building models by collecting a set of photos that densely cover the target area from many different viewpoints.


\subsection{Photo Crowdsourcing}
Image-based 3D reconstruction requires to collect sufficient and high quality fresh photos that best cover the reconstruction target area.
Photo crowdsourcing~\cite{8056963} (also known as mobile crowd photography~\cite{Guo2016, 8485969}) can be regarded as an ideal tool for photo collection, which requires large amounts of participants to contribute their photos via rich built-in sensors of their hand-held devices.
To guarantee good performance of photo crowdsourcing results, it is essential to select a proper set of photos.
Authors in~\cite{8056963} propose a photo selection mechanism for photo crowdsourcing.
The main objective is to select a set of photos to best cover the target area under limited network resources.
To this end, they first define photo utility based on its metadata (i.e., photo properties, such as location, orientation and field of view), in order to measure how well a target area is covered by a given set of photos.
Then, they design an efficient photo selection algorithm that can select photos with the largest utility under a resource constraint.
Note that the photo selection process is based on metadata instead of real images, thus reducing photo transmission overhead.
Real-time crowdsourcing is another important issue for photo crowdsourcing system.
Authors in~\cite{8315500} propose an optimal participation and reporting decisions (OPRD) algorithm to support delay-sensitive mobile crowdsensing applications.
The objective is to jointly optimize participation and reporting decisions for mobile users, and maximize the reward for service provider.
In addition, they consider the data collection through cellular or Wi-Fi networks.


\begin{figure*}[t!]
    \centering
    \includegraphics[width=6.5in]{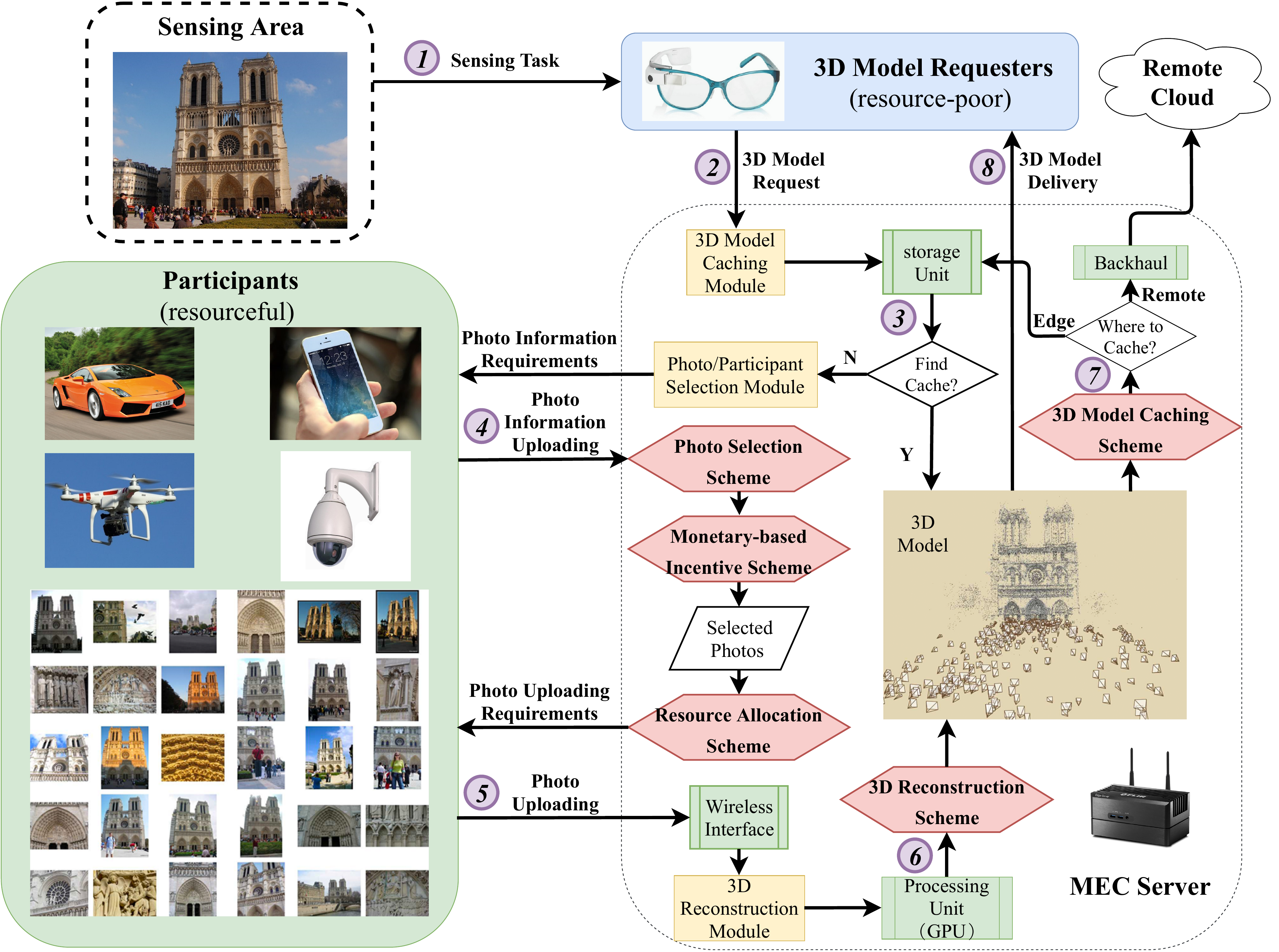}
    \caption{Proposed edge computing-based Photo Crowdsourcing (EC-PCS) framework.}
    \label{fig:EC-PCS}
\end{figure*}

\subsection{Edge Computing}
In the edge/fog computing architecture, distributed edge/fog servers are deployed close to end users that provide cloud-computing capabilities and IT services with ultra-low latency, high bandwidth, reliability and privacy protection.
Low latency task execution is a hot issue for researchers.
For example, authors in~\cite{7980005} study the latency-driven cooperative task computing in fog computing environments.
In their proposed framework, computational tasks can be jointly processed across multiple F-RAN nodes and near-range communications at network edge.
They propose a latency-driven cooperative task computing algorithm to minimize the task execution time, and characterize the tradeoff between communication and computing across multiple F-RAN nodes.
Deploying deep learning algorithms in edge/fog computing is another concern for researchers.
Authors in~\cite{8425301} develop robust mobile crowd sensing (RMCS) framework that integrates deep learning based data validation and edge computing based local processing.
The main objective is to provide robust data validation and local data processing for MCS by leverage MEC.
Authors in~\cite{10.1145/3219819.3220066} generalize the winning price model to incorporate the deep learning models with different distributions.
To this end, they propose an algorithm to learn from the historical bidding information, where the winning price are either observed or partially observed.

In this paper, we will present a novel photo crowdsourcing based 3D reconstruction method in 5G multi-access edge computing (MEC) environments.
MEC servers can quickly collect RGB images from ubiquitous 5G mobile and IoT devices in proximity that are equipped with built-in cameras.
Thus, omnidirectional 3D models can be reconstructed in real-time.
Compared with the traditional (e.g., centralized cloud based) photo crowdsourcing schemes (e.g.,~\cite{8056963,Guo2016, 8485969,8315500}), we focus on both real-time photo transmission on the network side and the selection of high quality fresh photos on the mobile user side, in order to realize real-time 3D reconstruction.
Indeed, our proposal also has the following benefits: i) real-time data aggregation and processing, since edge server is close to data sources, ii) reducing privacy leakage risks, since sensitive information is processed locally at edge servers, and iii) enabling location-aware real-time 3D services and applications.

\section{EC-PCS Overview}{\label{sec:System_Overview}}
In this section, we will first illustrate the framework of our edge computing-based photo crowdsourcing (EC-PCS).
Then, we present two illustrative examples for the EC-PCS applications in real-world scenario.
At last, the system model will be shown.

\subsection{System Framework}
The framework of our edge computing-based photo crowdsourcing (EC-PCS) is shown in Fig.~\ref{fig:EC-PCS}.
It consists of three main modules, which are photo/participant selection module, 3D reconstruction module and 3D model caching module, respectively.
At the beginning, a \textbf{mobile requester} has a sensing task (step 1), thus takes a photo of the objective(s) in target area from his viewpoint and sends it to a nearby MEC server (step 2).
The MEC server receives the photo, first feeds it to the 3D model caching module (details will be shown in Section~\ref{sec:Data_Caching_Module}).
The 3D model caching module detects the photo's feature by leveraging the scale-invariant feature transform (SIFT) algorithm~\cite{790410}, in order to find if the required 3D model is already cached (step 3) in the server.
If the required 3D model is i) cached locally, the MEC server sends the model to the requester immediately (step 8), or ii) not cached, the MEC server sends the request to the photo/participant selection module.
The photo/participant selection module (details will be shown in Section~\ref{sec:Participant_Selection_Module}) consists of a photo selection scheme, a monetary-based incentive scheme and a resource allocation scheme, responsible for select a set of proper photos from \textbf{edge participants} (i.e., the ubiquitous 5G mobile and IoT devices such as mobile phones, surveillance cameras and UAVs, as shown in Fig.~\ref{fig:EC-PCS}) to reconstruct the required 3D model.
To this end, the MEC server first broadcasts photo information uploading requests to all the edge participants (step 4), in order to collect the photos' information within the sensing area.
After receiving the information, MEC server selects an (near) optimal set of photos based on the photo selection scheme, and sends the photo uploading request to the edge participants that have selected photos.
Motivated by a monetary-based incentive scheme, the selected edge participants join the sensing task, and upload the selected photos to the MEC server (step 5).
Based on a resource allocation scheme, the MEC server receives the selected photos and feeds them to the 3D reconstruction module.
The 3D reconstruction module (details will be shown in Section~\ref{sec:3D_Reconstruction_Module}) responsible for reconstructing the required 3D model according to a 3D reconstruction scheme.
Note that the scheme in our implement later is based on the VisualSFM~\cite{VisualSFM}, and we also use GPU to speed up the reconstruction process.
After generating the required 3D model, the MEC server not only sends the model to the mobile requester (step 8), but also has to decide if the model is worth caching locally (step 7).
The decision making process is determined by a 3D model caching scheme, as shown in Fig.~\ref{fig:EC-PCS}.

Note that in the EC-PCS design later on, we allow the requester to set the target area coverage requirement as per its need.
Thus, when the number of edge participants is small, the requester can flexibly either increase its price or lower its coverage ratio requirement according to its application demand.
We should emphasize that when none or few fresh photos are available from the mobile and IoT devices in proximity and user's application is delay tolerant, the MEC can forward the user's request to the remote cloud, and then the operation turns into the traditional cloud-based photo crowdsourcing mode.
Thus, EC-PCS can be beneficial and complementary to existing approaches.

\subsection{Illustrative Examples}
\begin{figure}[t!]
    \centering
    \includegraphics[width=3.5in]{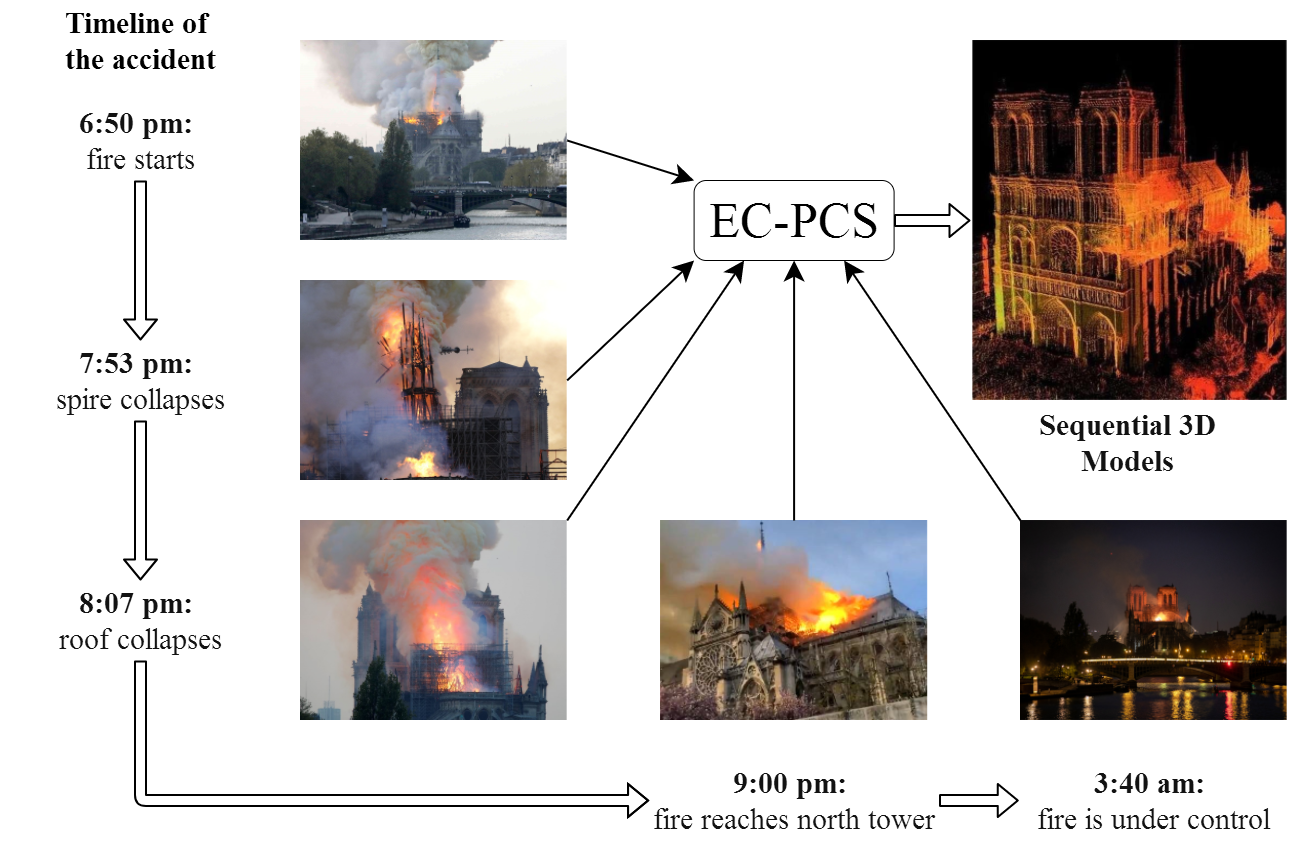}
    \caption{EC-PCS in the disaster monitoring \& recovery.}
    \label{fig:usecase2}
\end{figure}
\textit{Use Case 1: 5G Mobile Application Scenario}

For the emerging 5G mobile applications, an illustrative case study is the mobile augmented reality (AR) game that creates a 3D environment for the game players.
Assume that a mobile user Bob (mobile requester) is now playing a AR game using Google glass in the square of the Notre Dame de Paris, as shown in Fig. \ref{fig:EC-PCS}.
Note that running such mobile application requires the 3D model of objective (i.e., the Cathedral) in real-time.
Due to the limited computation and battery capacities of Google glass, Bob can not generate the required 3D model locally.
Moreover, Bob is in the front of the objective with limited field of vision, and impossible to obtain the whole 3D model of the Cathedral.
Thus, Bob has to send a 3D model request to a nearby MEC server by taking a photo of the objective.
If the MEC server can not find the 3D model of the Cathedral locally at its cache, the MEC server has to recruit ubiquitous 5G mobile and IoT devices (i.e., edge participants, such as smartphones, UAVs and surveillance cameras, as shown in Fig.~\ref{fig:EC-PCS}) to contribute their photos for the sensing task.
Note that the MEC server would collect tons of real-time captured photos if all the edge participants upload all their photos.
Thus, uploading all these photos is not necessary and inefficient, a representative subset is preferred.
One the other hand, the edge participants would willing to join the sensing task for the following two reasons: 1) inspired by a monetary-based incentive scheme, Bob will pay some money for the selected photos, 2) privacy protection for the selected photos, the photos will be processed at the edge side, instead of uploading their sensitive photos to the remote cloud server, thus reduce the risk of privacy leakage.
After receiving the selected photos, the MEC server performs 3D reconstruction and use GPU to speed up the process.
At last, the MEC server sends the required 3D model to Bob, and decides if the model worth caching locally for further reuse.
\\

\noindent \textit{Use Case 2: Disaster Monitoring \& Recovery Scenario}

Once again, we take the Notre Dame Cathedral as an example.
The famous building went up in flames on April 15, 2019, as shown in Fig. \ref{fig:usecase2}.
From the perspective of firefighters, they are urgent to know the latest fire behavior: where and how many are the ignition points, if structure of the building has been damaged, etc.
Thus, every moment's 3D model of the cathedral is required in real time.
On the one hand, from the perspective of tourists and residents around the cathedral, they would take photos all the time and share the photos on social networks.
In such disaster scenario, it makes sense to collect photos from the peoples and the surveillance cameras around the cathedral by leveraging our EC-PCS, in order to provide latest 3D model to the firefighters in real time.
Note that in the above photo crowdsourcing scenario, the freshness and viewpoint of each photo play key roles in photo selection.
To this end, our EC-PCS selects photos according to the photo's coverage, freshness, data size, resolution and the wireless channel states of their associated participants.

\subsection{System Model}
In this section, we will introduce the system model of the proposed EC-PCS.
For ease of reference, we list the key notations of our system model in TABLE.~\ref{table:keynotations}.

\begin{table}[t]
\caption{Key Notations}
\label{table:keynotations}
 \begin{tabular}{m{1.3in}<{}m{1.8in}<{}}\hline
 \toprule
\textbf{Symbol}&\textbf{Definition}\\
  \midrule
$\mathcal{M} = \{1, 2,..., M\}$                       & The set of edge devices (EDs). \\
$\mathcal{B}=\{B_{1},B_{2}...,B_{M}\}$                & The wireless resource allocation portfolio for the EDs. \\
$\mathcal{P}_{m}$=$\{P_{m,1}, P_{m,2},...,P_{m,N}\}$  & The photo set of edge participant (i.e., ED) $m$. \\
$\mathbb{SNR}_{m}$                                    & The Signal to Noise Ratio (SNR) between edge device $m$ and the MEC server. \\
$p_{m}^{u}$                                           & The transmit power of the edge device $m$. \\
$d_{m}$                                             & The distance between edge device $m$ and the MEC server. \\
$L_{m}$                                               & The current location for edge participant $m$. \\
$D_{m,n}$                                             & The data size (in MB) of photo $P_{m,n}$. \\
$t_{m,n}$, $l_{m,n}$                                  & When and where the photo $P_{m,n}$ is taken.\\
$\overrightarrow{d_{m,n}}$                            & The orientation (i.e., viewing direction) of camera when the photo $P_{m,n}$ is taken.\\
$\varphi_{m,n}$                                       & The field of view of photo $P_{m,n}$.\\
$r_{m,n}$                                             & How far the camera of edge participant $m$ can see.\\
$G_{m,n}$                                             & The price of photo $P_{m,n}$.\\
$\rho_{m,n}$                                          & The resolution of photo $P_{m,n}$.\\
$\mathcal{A}^{Tar}$                                   & The target area of a 3D reconstruction task.\\
  \bottomrule
 \end{tabular}
\end{table}

\subsubsection{Network Model}
The EC-PCS framework is composed of one MEC server and a set of edge devices $\mathcal{M} = \{0, 1, 2,..., M\}$ (i.e., 5G mobile and IoT devices) in the sensing area.
Without loss of generality, let $m=0$ $(m\in\mathcal{M})$ denote the mobile requester, and $m=1,2,\cdots, M $ $(m\in\mathcal{M})$ represent the $M$ edge participants.
The edge devices can establish 5G cellular connections with the MEC server, and the connections are based on the orthogonal frequency division multiple-access (OFDMA) \cite{OFDMA}.
Since the edge devices within the coverage of MEC server occupy different frequency subcarriers, these devices would not interfere with each other under OFDMA.
In this paper, a total bandwidth of $B$ is available to the edge devices for the sensing task.
Moreover, let $B_{m}$ $(m\in\mathcal{M})$ denote the bandwidth that is allocated to edge device $m$, and $\mathcal{B}=\{B_{1},B_{2}...,B_{M}\}$ represents the resource allocation portfolio of all the edge devices.
Thus, we have $\sum_{m=1}^{M}B_{m}=B$.

Based on the above network model, the maximum uplink rate in (bps), achievable for an edge device $m$ $(m\in \mathcal{M})$ during photo uploading procedure, over an additive white Gaussian noise (AWGN) channel, can be expressed as follows:

\begin{equation}{\label{eq:sinr}}
r_{m}^{ul}=B_{m}\log_{2}\left(1+\mathbb{SNR}_{m}\right),
\end{equation}
where $\mathbb{SNR}_{m}$ denotes the Signal to Noise Ratio (SNR) between edge device $m$ and the MEC server.
More specifically,
\begin{equation}{\label{eq:price}}
\begin{aligned}
\mathbb{SNR}_{m}=\frac{p_{m}^{u}|h_{m}^{ul}|^{2}}{\Gamma(g_{ul})d_{m}^{\beta}N_{0}},
\end{aligned}
\end{equation}
where $p_{m}^{u}$ denotes the transmit power of the edge device $m$, $d_{m}$ is the distance between edge device $m$ and the MEC server, and $N_{0}$ represents the noise power.
On the other hand, $h_{m}^{ul}$ denotes the channel fading coefficient (in Rayleigh-fading environment) between edge device $m$ and the MEC server, with $\beta$ being the path loss exponent.
Lastly, the function $\Gamma(\mathbb{BER})=-\frac{2\mathrm{ln}(5\mathbb{BER})}{3}$ is the minimum SNR threshold that achieves the desired target bit error rate (BER), and $g_{ul}$ refers to the target uplink BER.
Notice that we assume that these parameters are not controllable, similar to the assumptions made in \cite{DREAM}.

\subsubsection{Photo properties}
\begin{figure}[t!]
    \centering
    \includegraphics[width=3in]{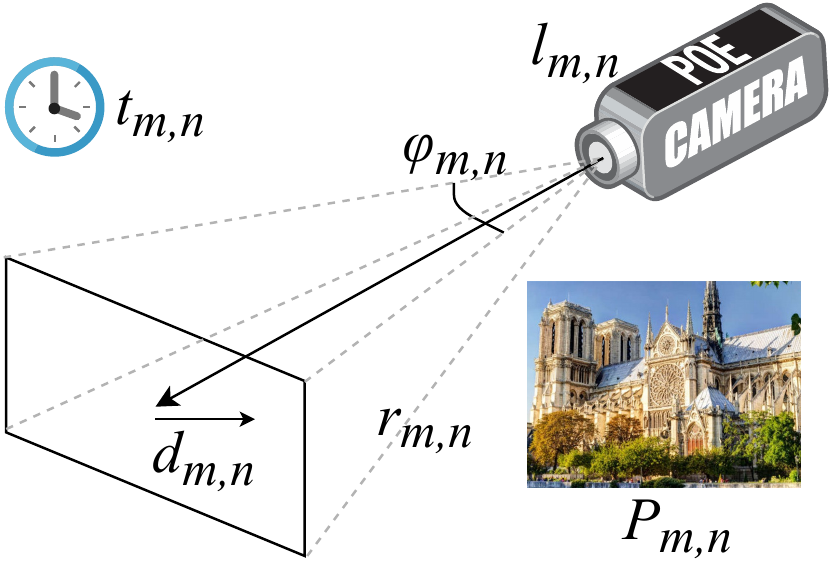}
    \caption{Photo Coverage area for a camera.}
    \label{fig:camera}
\end{figure}

Assume that each edge device $m$ can store at most $N$ photos.
Let $\mathcal{P}_{m}$=$\{P_{m,1}, P_{m,2},...,P_{m,N}\}$ represents the photo set of edge participant $m$.
We adopt a parameter tuple $I_{m,n}$=$\{L_{m}, D_{m,n}, t_{m,n}, l_{m,n}, \varphi_{m,n}, \overrightarrow{d_{m,n}}, r_{m,n}\}$ $(I_{m,n}\in\mathcal{I}_{m})$ to characterize the properties for the photo $P_{m,n}$, where $L_{m}$ is the current location for edge participant $m$, $D_{m,n}$ represents the data size (in MB) of photo $P_{m,n}$.
As shown in Fig.~\ref{fig:camera}, $t_{m,n}$, $l_{m,n}$, $\varphi_{m,n}$, $\overrightarrow{d_{m,n}}$ and $r_{m,n}$ are metadata~\cite{8056963} of photo $P_{m,n}$, where $t_{m,n}$ and $l_{m,n}$ denote when and where the photo is taken.
Vector $\overrightarrow{d_{m,n}}$ is the orientation (i.e., viewing direction) of camera when the photo is taken.
$\varphi_{m,n}$ represents field of view of photo $P_{m,n}$, such angle specifying how wide the camera can see.
Range $r_{m,n}$ specifies how far the camera can see.
Note that the above photo properties can be obtained from the APIs and built-in sensors of most mobile devices~\cite{8056963}.
The properties can also be estimated by VisualSFM algorithm~\cite{VisualSFM}.

\section{Photo/Participant Selection Module}{\label{sec:Participant_Selection_Module}}
As stated earlier, the main function of photo/participant selection module is to select a representative set of photos for image-based 3D reconstruction by leveraging photo crowdsourcing.
In the EC-PCS, the MEC server recruits a large number of participants to collect photos.
Moreover, the participants may have tons of real-time captured photos for a target area and most of the photos may contain duplicated information.
It will be inefficient and time consuming to upload all the photos to the MEC server.
Thus, it makes sense to balance the computation overhead (i.e., 3D reconstruction) and communication cost (i.e., photo uploading).
To this end, we will present a monetary-based incentive scheme, a photo selection scheme and a resource allocation scheme for the photo/participant selection module.

\subsection{Monetary-based Incentive Scheme}
In the photo crowdsourcing process, the MEC server would collect photos from multiple edge participants.
However, edge participants are reluctant to share their photos due to the lack of sufficient incentives.
On the one hand, participating in the photo crowdsourcing task may incur additional costs for edge participants, such as computation, communication, and energy overhead on their devices.
On the other hand, the collected photos usually contain sensitive information, such as location information and face information.
Therefore, it is conceivable that edge participants will not participate in this photo crowdsourcing task, unless they are properly motivated.

Paying for selected photos in photo crowdsourcing is the most intuitive incentive.
Edge participants who are willing to make some money can sell their photos for the crowdsourcing task.
Thus, it is critical to pricing the photos.
Note that pricing the photos (also be known as defining the photo utility) is a key issue in photo crowdsourcing schemes~\cite{8056963,8485969}, and is decided by the main objective of the photo crowdsourcing scheme.
Since the main objective of the EC-PCS is to select a proper set of photos that can minimize the 3D reconstruction.
The photo quality (i.e., resolution and freshness) and photo transmission overhead (i.e., photo data size and wireless channel state) should be jointly considered.
Thus, we adopt a heuristic pricing scheme of photo $P_{m,n}$ $(m\in \mathcal{M}, n=1,2,...,	N)$ as follows:
\begin{equation}{\label{eq:price}}
\begin{aligned}
G_{m,n}=\omega\cdot\frac{\rho_{m,n}(t_{0}-t_{m,n})}{D_{m,n}\,\mathbb{SNR}_{m}},
\end{aligned}
\end{equation}
where $G_{m,n}$ and $\rho_{m,n}$ represent the price and resolution of photo $P_{m,n}$, $\omega$ is a price scaling factor that can be set by the task requester (e.g., as per its budget and coverage requirement).
$t_{0}$ is the current system time, thus $t_{0}-t_{m,n}$ represents the freshness of photo $P_{m,n}$.

Such scheme quantifies the price for the photos, where photos' freshness, resolution and data size, as well as their associated edge participants' wireless channel states are jointly considered.
The main objective of this pricing strategy is to evaluate each photo's contribution for the photo selection scheme in the next step.
Note that our primary focus of this study is the system performance optimization of EC-PCS via efficient photo selection and network resource allocation, and other pricing mechanisms for incentivizing participants can also be applied in our framework.

\subsection{Photo Selection Scheme}
During photo crowdsourcing process, photo selection is performed by the MEC server to obtain a photo set with certain coverage considering photo uploading constraints.
Note that the edge participants (e.g., UAVs and Surveillance Cameras) usually contain tons of real-time captured photos with varied distances to the objective, and different shooting angles in the sensing area.
Due to the limited computation and communication resources, it is inefficient and unnecessary to upload all the photos of the edge participants.
In this work, we consider a low-overhead photo information uploading scheme, similar to~\cite{8056963}.
All the edge participants need only to upload their photo properties $\mathcal{I}_{m}$ to the MEC server, in order to: i) locate the target area for the objective, and ii) provide photo information for a photo selection mechanism.

\begin{figure}[t!]
    \centering
    \includegraphics[width=3.4in]{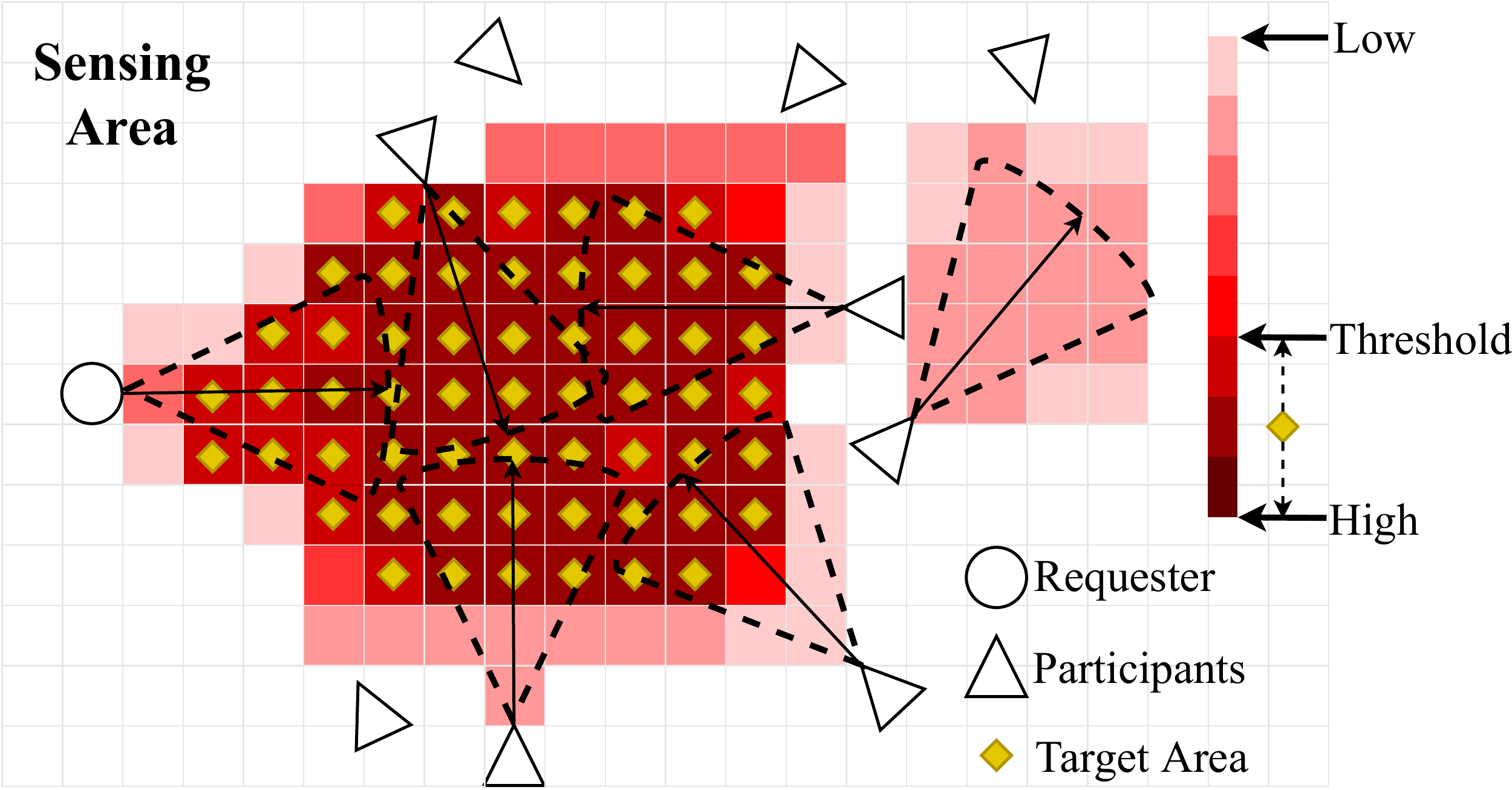}
    \caption{Locating target area by crowdsourced photos.}
    \label{fig:locating_target_area}
\end{figure}

\subsubsection{Locating the Target Area}
Although the photos from different edge participants have different locations and different shooting angles, they all prone to shoot towards the target area.
For example, multiple tourists in the Notre Dame square prone to shoot photos of the cathedral with their hand-held devices.
Therefore, the location of target area can be estimated by calculating the most overlapped areas of crowdsourced photos~\cite{7524481}, as shown in Fig.~\ref{fig:locating_target_area}.
In this paper, we extend the 2D InstantSense scenario~\cite{7524481} into our 3D sensing scenarios, where edge participants collaborate to cover different aspects of the target area.
Assume that the 3D sensing space can be divided into $G^{3}$ grids where the length, width and height of each grid are all set to be $1$m.
Let a set $\mathcal{G}$=$\{g_{i,j,k}\}$ $(i,j,k=1,2,...,G)$ denote the grid set of the sensing space, and $a_{i,j,k}$ represent binary indicator variables of $g_{i,j,k}$.
$a_{i,j,k}=1$ indicates grid $g_{i,j,k}$ is in the target area, and $a_{i,j,k}=0$, otherwise.
Since the shooting direction of each photo has the maximum probability of passing through the target area, the grids have different number of times covered by photos.
Given a pre-define threshold of number of coverage, the set of grids $\mathcal{A}^{Tar}$ that covered by the target area can be estimated as follows:
\begin{equation}{\label{eq:price}}
\begin{aligned}
\mathcal{A}^{Tar}&=\{g_{i,j,k}\},\\
a_{i,j,k}&=1, \\
i,j,k&=1,2,...,G.\\
\end{aligned}
\end{equation}
When a target area $\mathcal{A}^{Tar}$ is determined, the aspects of the target area that is covered by a photo $P_{m,n}$ can be obtained as $\mathcal{A}^{Tar}_{m,n}$.
Note that the sets $\mathcal{A}^{Tar}_{m,n}$ $(m=1,2,...,M; n=1,2,...,N)$ are the subsets of $\mathcal{A}^{Tar}$.

\subsubsection{Cost-aware Photo Selection}
Since the performance of 3D reconstruction relies heavily on the quality of selected photos, we should select photos that best cover the target area.
Note that the selected photos can be: i) the photos just captured by edge participants, or ii)  relevant photos stored in the participants' albums.
Thus, we must consider the freshness of the photos, in order to depict the most recent state of the target area.
Moreover, edge participants are reluctant to upload their personal photos.
The mobile requester has to pay some money to the participants.
The challenge faced in this process is that the selected photo set should satisfy the required target area coverage with the minimum monetary cost.
To this end, the optimization problem for the photo selection scheme can be summarized as follows:
\begin{equation}
\begin{aligned}
\label{eq:optimal_photo}
\mathcal{P}^{*}=\operatorname*{argmin} &\sum\limits_{P_{m,n}\in \mathcal{P}^{*}} G_{m,n},\\
\mathrm{s.t. }\quad&|\cup_{m\in\mathcal{M},P_{m,n}\in\mathcal{P}_{m}}\mathcal{A}^{Tar}_{m,n}|\geq \eta\cdot|\mathcal{A}^{Tar}|, \\
\end{aligned}
\end{equation}
where $\mathcal{P}^{*}$ is the optimal (i.e., selected) set of photos, $\eta\in(0,1]$ represents the coverage ratio factor, $|\mathcal{A}|$ denotes the number of elements in set $\mathcal{A}$.
Note that the constraint in~(\ref{eq:optimal_photo}) guarantees the selected photos must cover at least $\eta$ of the target area.

\begin{thm}
\label{prop:nphard}
The cost-aware photo selection problem in (\ref{eq:optimal_photo}) is NP-hard.
\end{thm}
\noindent The detailed proof is provided in \textbf{Proof A} of this paper.

The key idea is to show that our problem in (\ref{eq:optimal_photo}) can be reduced from a vertex cover problem~\cite{VCP}, which is also NP-hard.
Thus, we utilize an approximation algorithm based on a greedy strategy~\cite{Wolsey1982} to obtain a near-optimal set of photos.
Our greedy-based algorithm is listed in \textbf{Algorithm~\ref{alg:greedy}}.

The main principle of Algorithm~\ref{alg:greedy} is to iteratively find the best photo which has the maximum value of number of covered grids divided by monetary cost.
Then, add the photo to the set of selected photos (i.e., $\mathcal{P}^{*}$).
The algorithm stops when the minimum coverage requirement $\eta\cdot|\mathcal{A}^{Tar}|$ is achieved.
The time complexity for the algorithm is $\mathcal{O}(n^{2})$.

We can theoretically characterize the approximation ratio (i.e., the ratio of the solution by the proposed algorithm over the optimal solution in the worst case) of the Algorithm 1 as follows:

\begin{algorithm}[t]
\caption{The greedy-based photo selection algorithm.}
\label{alg:greedy}
  \begin{algorithmic}[1]
    \Require
      The set of photos \{$P_{m,n}$\}, $(m\in\mathcal{M}, n=1,2,...,N)$;
      The price of photo $P_{m,n}$, $G_{m,n}$;
    \Ensure
      A set of selected photos, $\mathcal{P}^{*}$;
    \Function{Greedy}{$G_{m,n}$, $P_{m,n}$}
    \State Initialize: $flag_{m,n}$=false, U=$\mathcal{A}^{Tar}$ from all $P_{m,n}$
    \While{$c>0$}
        \For{$P_{m,c}$ \textbf{in} $P_{m,n}$}
            \If{$flag_{m,c}=$false}
             \State run count\_covered\_grids($P_{m,c}$, $U$)
             \State weight=covered\_grids/$G_{m,c}$
             \State choose the max weight $P_{m,max}$
             \State $U-$points in $P_{m,max}$
             \State $flag_{m,max}=$true
            \EndIf
        \EndFor
    \EndWhile
    \State filter the photos where $flag_{m,n}$=false
    \State \Return $\mathcal{P}^{*}$
    \EndFunction
  \end{algorithmic}
\end{algorithm}

\begin{thm}
\label{prop:approximation}
Algorithm~\ref{alg:greedy} for the cost-aware photo selection problem achieves an approximation ratio of $F_{\lceil \eta\cdot|\mathcal{A}^{Tar}| \rceil}$, where $F_{n}=\sum\limits_{i=1}^{n}\frac{1}{i}$.
\end{thm}
The detailed proof is provided in \textbf{Proof B} of this paper.
Numerical results in Section~\ref{sec:performance_evaluation} corroborate the superior performance of Algorithm 1.

\subsection{Network Resource Allocation Scheme}
Once the set $\mathcal{P}^{*}$ is determined, the edge participant $m$ $(m\in\mathcal{M})$ needs to upload a set $\{{P}_{m,n}, {P}_{m,n}\in\mathcal{P}^{*}\}$ of photos in its local album.
As a result, the uploading delay $T_{m}^{u}$ for edge participant $m$ to upload its selected photos can be expressed as follows:
\begin{equation}
\label{eq:uploading_delay_single}
T_{m}^{u}(\mathcal{B})=\frac{\sum_{P_{m,n}\in \mathcal{P}^{*}} D_{m,n}}{B_{m}\log_{2}\left(1+\mathbb{SNR}_{m}\right)}.
\end{equation}
To achieve real-time 3D reconstruction, we have to minimize the global photo uploading delay.
With the expression of $T_{m}^{u}$, the global photo uploading delay of the EC-PCS framework is given by
$\mathrm{max}\left\{T_{1}^{u}(\mathcal{B}),\cdots,T_{M}^{u}(\mathcal{B})\right\}$.
Therefore, to achieve real-time 3D reconstruction, the key is to minimize the uploading delay by designing the bandwidth $\mathcal{B}$:
\begin{equation}
\begin{aligned}
\label{eq:uploading_delay}
\mathcal{B}^{*}=\operatorname*{argmin}_{\mathcal{B}}\quad&\mathrm{max}\left\{T_{1}^{u}(\mathcal{B}),T_{2}^{u}(\mathcal{B}),\cdots,T_{M}^{u}(\mathcal{B})\right\},\\
\mathrm{s.t.}\quad&\sum_{m=1}^{M}B_{m}=B.\\
\end{aligned}
\end{equation}
With introduction of a slack variable $t$ such that $t\geq\mathrm{max}\left\{T_{1}^{u}(\mathcal{B}),T_{2}^{u}(\mathcal{B}),\cdots,T_{M}^{u}(\mathcal{B})\right\}$, the min-max resource allocation problem in (\ref{eq:uploading_delay}) can be equivalently converted into:
\begin{equation}
\begin{aligned}
\label{eq:uploading_delay_convert}
\min\limits_{\mathcal{B},t}\quad&t,\\
\mathrm{s.t.}\quad &h_{i}(\mathcal{B},t)\leq 0, \quad i=1,2,...,M,\\
&g(\mathcal{B})=0,
\end{aligned}
\end{equation}
where $h_{i}(\mathcal{B},t)=T_{i}^{u}(\mathcal{B})-t$ and $g(\mathcal{B})=\sum_{m=1}^{M}B_{m}-B$.
By deriving the KKT conditions of problem (\ref{eq:uploading_delay_convert}), we can obtain the closed-form solution for the original problem as follows:
\begin{thm}
\label{prop:convex}
The optimal solution to the resource allocation problem in (\ref{eq:uploading_delay}) is $\mathcal{B}^{*}=\{B_{1}^{*},B_{2}^{*},...,B_{M}^{*}\}$, where
\begin{align}
B_{m}^{*}=\dfrac{B\cdot\dfrac{\sum_{P_{m,n}\in \mathcal{P}^{*}} D_{m,n}}{\log_{2}(1+\mathbb{SNR}_{m})}}{\sum_{m=1}^{M}\left(\dfrac{\sum_{P_{m,n}\in \mathcal{P}^{*}} D_{m,n}}{\log_{2}(1+\mathbb{SNR}_{m})}\right)}.
\end{align}
\end{thm}

The detailed proof is provided in \textbf{Proof C} of this paper.
Theorem~3 indicates that the bandwidth allocation is not only related to channel conditions, but also determined by the photo selection.
This is in contrast to traditional communication systems where channel quality is the only consideration.
The allocated bandwidth to a particular user is linearly proportional to its photo sizes and number of selected photos.
The bandwidth is also inversely proportional to the logarithm of SNR.

\section{3D Reconstruction and 3D Model Caching Modules}{\label{sec:3D_Reconstruction_Module}}
\subsection{3D Reconstruction Module}
When the MEC server receives the set of selected photos $\mathcal{P}^{*}$, a 3D reconstruction scheme is performed in the MEC server to reconstruct the required 3D model.
One of the widely used technique is called Structure from Motion (SFM)~\cite{SfM}, which can generate a 3D model in the state of motion.
The main principle of the SFM is to search for the same feature points from different photos by analyzing the relative and absolute positions of the field image.
The location and orientation of the camera can also be estimated by SFM.
Based on the SFM, a number of software was developed, such as Meshlab~\cite{Meshlab} and VisualSFM~\cite{VisualSFM}.
In this paper, we use the more popular and free option VisualSFM as the 3D reconstruction scheme for our EC-PCS framework.

\subsection{3D Model Caching Module}{\label{sec:Data_Caching_Module}}
As shown in Fig.~\ref{fig:EC-PCS}, the 3D model caching module of our EC-PCS has two main functions:

i) Checking if the required 3D model has already cached in the MEC server (i.e., step 3 in Fig.~\ref{fig:EC-PCS}).
After receiving the photo (i.e., task request) from the mobile requester, the MEC server detects and matches keypoints of the photo by leveraging Scale-Invariant Feature Transform (SIFT)~\cite{790410}.
For the computation part, the computational complexity of algorithm SIFT for 3D reconstruction is $\mathcal{O}(ij+k)$~\cite{6034893}, where $i$ and $j$ are the width and height of an image, $k$ represents the number of keypoints of the image, which can be well supported by the MEC server.
ii) Deciding if the reconstructed 3D model is worth caching in the MEC server (i.e., step 7 in Fig.~\ref{fig:EC-PCS}).
Due to the limited storage capacity of the MEC server, it is impossible to cache all the reconstructed 3D models of the sensing area.
Thus, it makes sense to cache the most popular 3D models that are likely to be reused in the future.
In this paper, we consider a popularity-based caching policy \cite{Zipf} that the MEC server store 3D models based on their highest popularity until the storage is achieved.
The 3D model's popularity distribution conditioned on the history that mobile requesters make 3D model requests.

\begin{figure}[t]
\centering
\subfigure[Old school gate of Tsinghua University.]{
\begin{minipage}[t]{0.5\linewidth}
\centering
\includegraphics[width=1.7in]{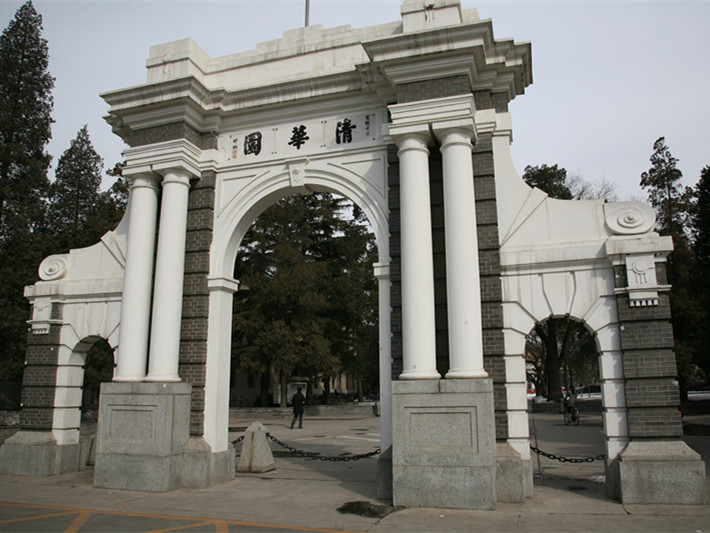}
\end{minipage}%
}%
\subfigure[Zhantan Temple.]{
\begin{minipage}[t]{0.5\linewidth}
\centering
\includegraphics[width=1.7in]{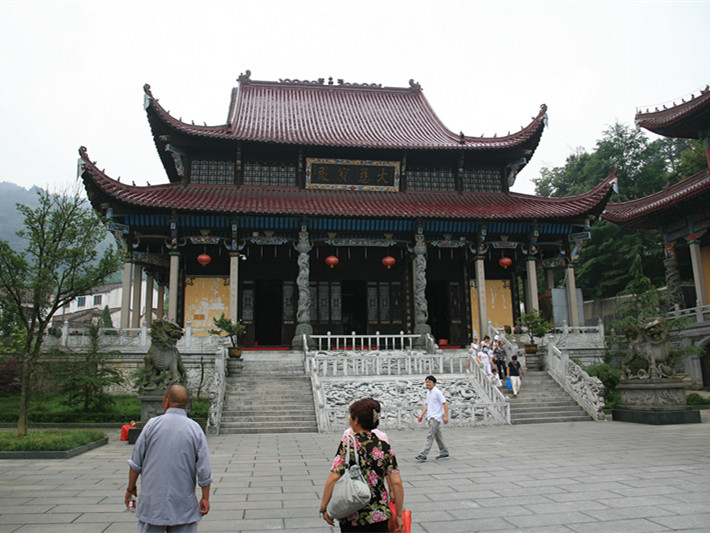}
\end{minipage}%
}
\centering
\caption{Two real-world photo datasets~\cite{data_set}.}
\label{fig:Data_Set}%
\end{figure}

\begin{footnotesize}
\begin{table}[!t]
\caption{Data Set Parameters.}
\label{tab:Data_Set_Parameters}
\centering
\renewcommand\arraystretch{1.2}
\begin{tabular}{|m{27mm}<{\centering}|m{20mm}<{\centering}|m{20mm}<{\centering}|m{20mm}<{\centering}|}\hline
Data Set &Data Size&\#Photos\\\hline
GATE & $340$MB  &   $68$ \\\hline
TEMPLE& $719$MB                    &   $141$        \\\hline
\end{tabular}
\end{table}
\end{footnotesize}

\section{Performance Evaluation}{\label{sec:performance_evaluation}}
In this section, we evaluate the performance of the proposed EC-PCS framework.
First of all, we describe our experimental environments.
Then, we examine the effectiveness of the photo selection scheme and network resource allocation scheme, respectively.

\subsection{Experimental Setup}
In our simulations, we use a desktop as the MEC server, which has a GPU of Nvidia Geforce GTX 1050.
The CPU is Intel i7-7700 with 8G memory.
Software environment we utilized are i) MATLAB R2018b to simulate the edge computing network environments and ii) VisualSFM~\cite{VisualSFM} to perform 3D reconstruction.
We evaluate the performance of our EC-PCS framework based on two real-world photo datasets: old school gate of Tsinghua University (GATE, as shown in Fig.~\ref{fig:Data_Set} (a)) and Zhantan Temple (TEMPLE, as shown in Fig.~\ref{fig:Data_Set} (b))~\cite{data_set}.
The main difference between the two datasets is the different number of objectives in the target area.
Note that GATE has only one objective (i.e., the gate) in the target area, whereas TEMPLE has three objectives (i.e., temples).
Thus, TEMPLE has a more complex 3D model than GATE, results in higher reconstruction cost.
TABLE~\ref{tab:Data_Set_Parameters} shows the parameters of the two datasets, and TABLE~\ref{tab:Network_Parameters} reports the parameters of simulated network environments.

\begin{footnotesize}
\begin{table}[!t]
\caption{Network Parameters.}
\label{tab:Network_Parameters}
\centering
\renewcommand\arraystretch{1.2}
\begin{tabular}{|m{17mm}<{\centering}|m{17mm}<{\centering}||m{17mm}<{\centering}|m{17mm}<{\centering}|}\hline
Parameter&Value&Parameter&Value\\\hline
$M$                 &$10$                  &$N$                 &$20$             \\\hline
$\text{SNR}_{m}$   &$(0,30]$ dB           & $t_{0}-t_{m,n}$    &$(0,10]$ min     \\\hline
$B$                 & $10$ MHz             & $\omega$            &$0.1$            \\\hline
\end{tabular}
\end{table}
\end{footnotesize}

\subsection{Performance for Photo Selection Scheme}
We implement our photo selection scheme for EC-PCS, and compare it with respect to two benchmark related policies, namely:

\begin{itemize}
\item \emph{Random Photo Selection Scheme (RPSS):} \\
Photos are randomly selected from the raw data set.
\item \emph{Clustering-based Photo Selection Scheme (CPSS):} \\
As introduced by~\cite{8485969}, photos are clustered into groups, and the scheme selects one photo from each cluster.
\end{itemize}

\begin{figure}[t]
\centering
\subfigure[$\eta=80\%$]{
\begin{minipage}[t]{0.5\linewidth}
\centering
\includegraphics[width=1.7in]{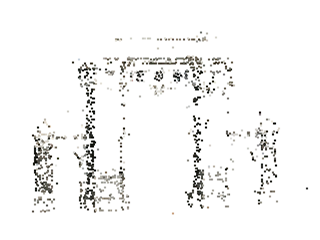}
\end{minipage}%
}%
\subfigure[$\eta=95\%$]{
\begin{minipage}[t]{0.5\linewidth}
\centering
\includegraphics[width=1.7in]{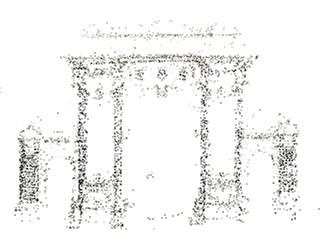}
\end{minipage}%
}%
\quad
\subfigure[$\eta=99.5\%$]{
\begin{minipage}[t]{0.5\linewidth}
\centering
\includegraphics[width=1.7in]{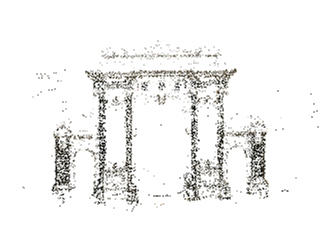}
\end{minipage}
}%
\subfigure[$\eta=100\%$]{
\begin{minipage}[t]{0.5\linewidth}
\centering
\includegraphics[width=1.7in]{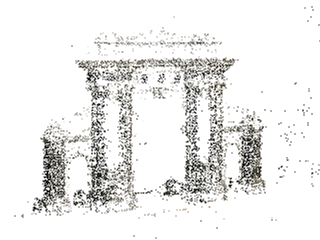}
\end{minipage}
}%
\centering
\caption{Evaluation results for the GATE dataset.}
\label{fig:3D_qinghua}
\end{figure}

\begin{figure}[t]
\centering
\subfigure[$\eta=80\%$]{
\begin{minipage}[t]{0.5\linewidth}
\centering
\includegraphics[width=1.7in]{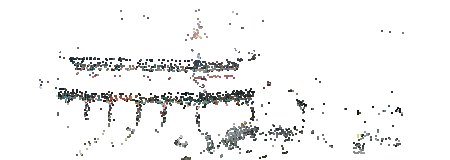}
\end{minipage}%
}%
\subfigure[$\eta=95\%$]{
\begin{minipage}[t]{0.5\linewidth}
\centering
\includegraphics[width=1.7in]{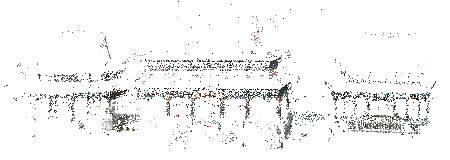}
\end{minipage}%
}%
\quad
\subfigure[$\eta=99.5\%$]{
\begin{minipage}[t]{0.5\linewidth}
\centering
\includegraphics[width=1.7in]{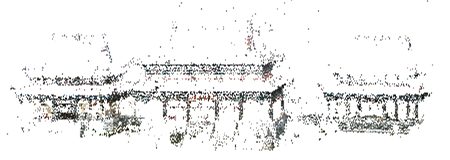}
\end{minipage}
}%
\subfigure[$\eta=100\%$]{
\begin{minipage}[t]{0.5\linewidth}
\centering
\includegraphics[width=1.7in]{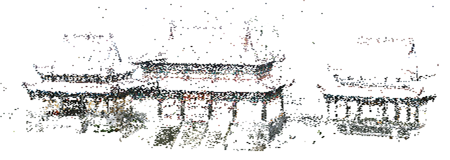}
\end{minipage}
}%
\centering
\caption{Evaluation results for the TEMPLE dataset.}
\label{fig:3D_zhantan}
\end{figure}

The main objective of our photo selection scheme is to select a proper set of photos that can: i) minimize the monetary cost, and ii) achieve a certain target area coverage.
Fig.~\ref{fig:3D_qinghua} and Fig.~\ref{fig:3D_zhantan} illustrate the quality of reconstructed 3D models versus different coverage ratio parameter $\eta$.
TABLE.~\ref{tab:GATE_result} and TABLE.~\ref{tab:TEMPLE_result} report the corresponding reconstruction costs, including the number of selected photos and total monetary cost, respectively.

From the experimental results, we can find that achieving a $100\%$ coverage is not necessary and expensive.
Note that Fig.~\ref{fig:3D_qinghua} (c) (i.e., the 3D model whose $\eta=99.5\%$) and Fig.~\ref{fig:3D_qinghua} (d) (i.e., the 3D model whose $\eta=100\%$) have a very similar reconstruction quality.
However, the method in Fig.~\ref{fig:3D_qinghua} (c) can reduce $25\%$ photo uploading and $36.1\%$ monetary cost, when compared with the method in Fig.~\ref{fig:3D_qinghua} (d).
Thus, it makes sense to set $\eta=99.5\%$ for real-world 3D reconstruction scenario.
The reason for the above phenomenon is that some of the selected photos have very limited contributions to the final 3D model.
Take the GATE as an example, we find that the last ten percent of the photos contribute less than 10 new keypoints for the final 3D model (the final 3D model consists of 20033 keypoints).
As a result, uploading the last $10\%$ of the photos to obtain less than the $0.5\%$ new keypoints is proven to be unnecessary.

Similar experimental results of TEMPLE are given in Fig.~\ref{fig:3D_zhantan} and TABLE~\ref{tab:TEMPLE_result}.
We can reduce $29.78\%$ photo uploading and $52.30\%$ monetary cost by setting $\eta=99.5\%$, and guarantee the quality of 3D model at the same time.

\begin{figure}[t]
\centering
\subfigure[GATE]{
\begin{minipage}[t]{0.5\linewidth}
\centering
\includegraphics[width=1.6in]{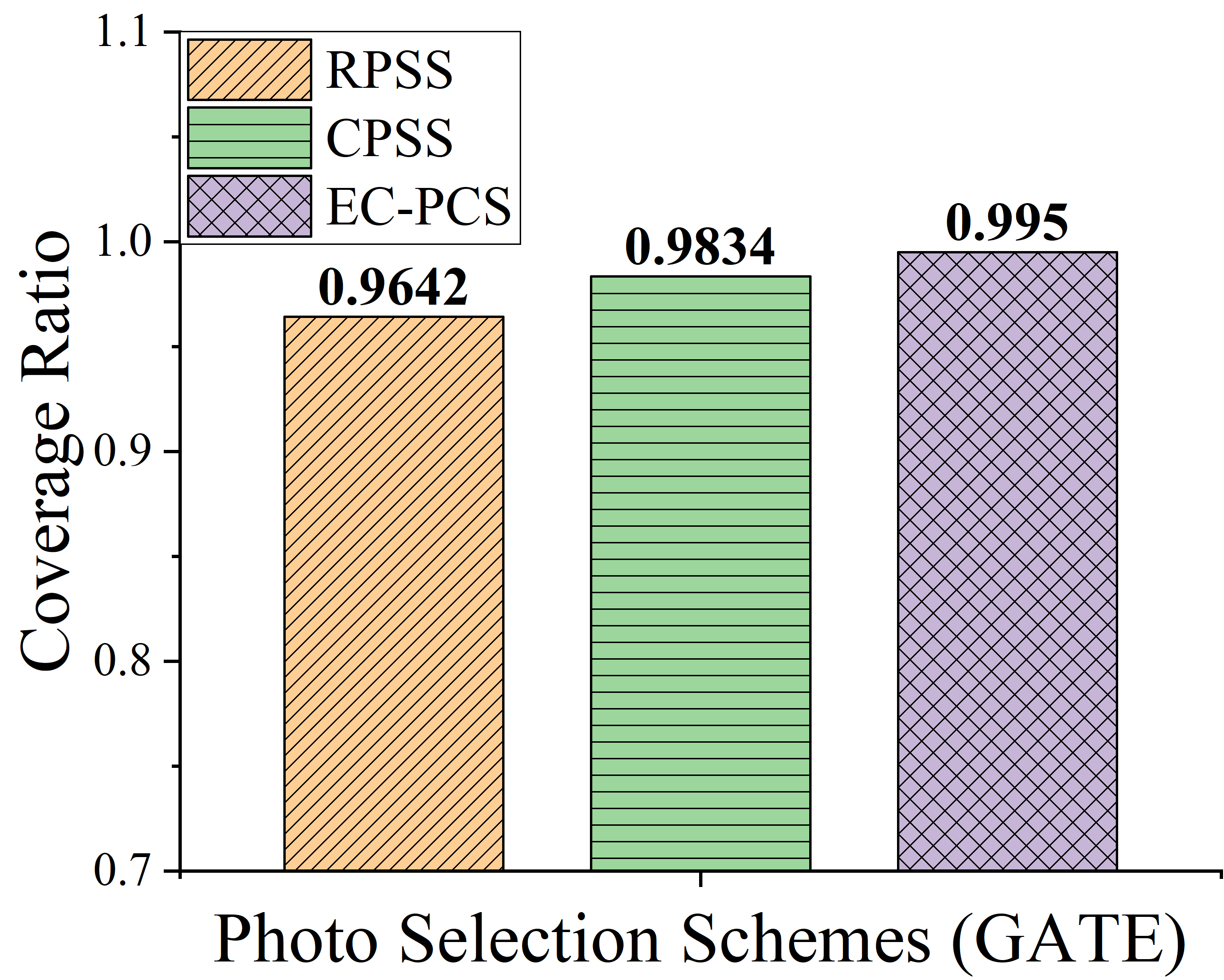}
\end{minipage}%
}%
\subfigure[TEMPLE]{
\begin{minipage}[t]{0.5\linewidth}
\centering
\includegraphics[width=1.6in]{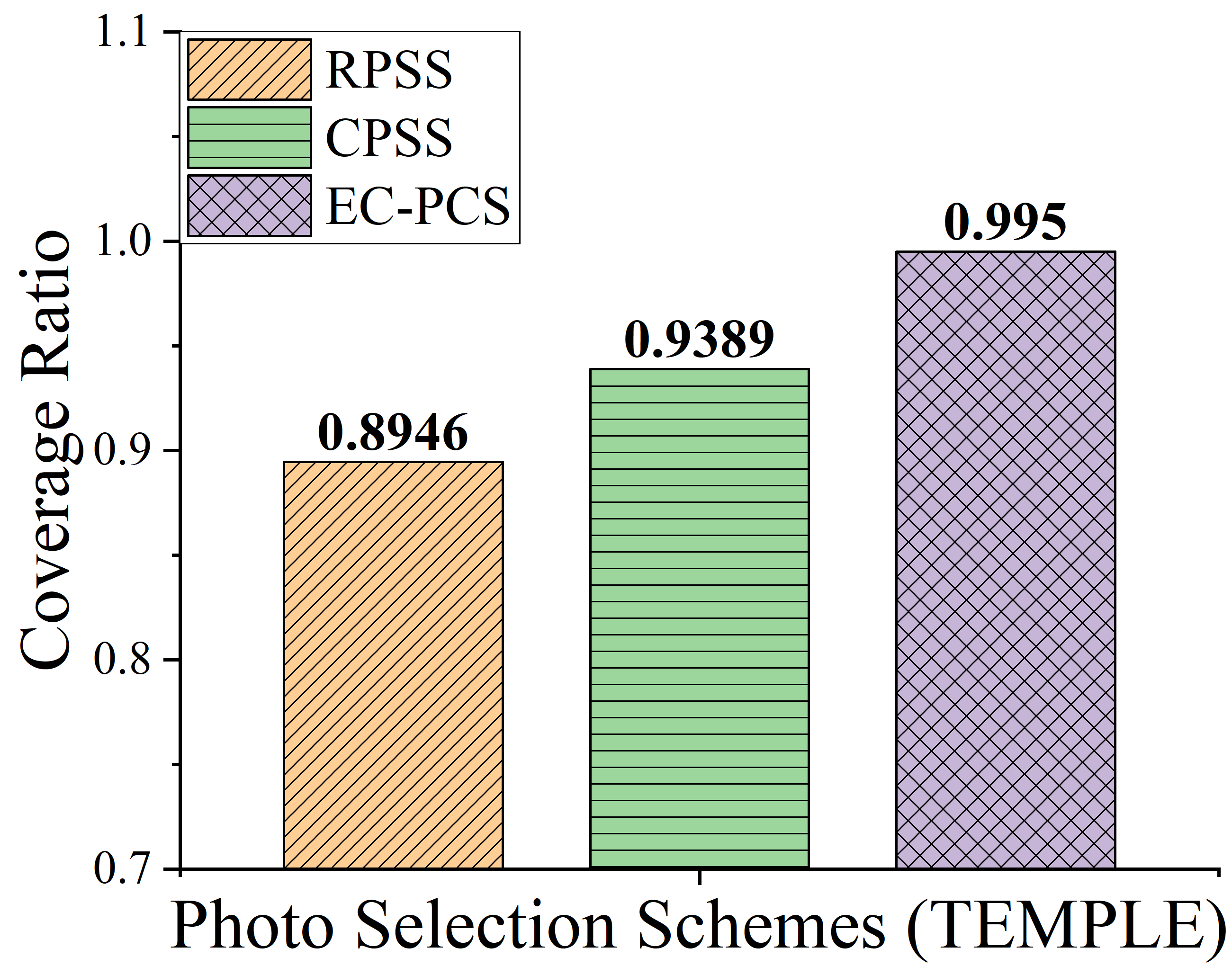}
\end{minipage}%
}%
\caption{Photo selection schemes vs. target area coverage ratio.}
\label{fig:3D-coveage}
\end{figure}

Fig.~\ref{fig:3D-coveage} shows the target area coverage ratio for different photo selection schemes.
In order to ensure fairness, we randomly select $50$ and $100$ photos from GATE and TEMPLE for RPSS, same as our EC-PCS mothod.
Note that our photo selection scheme for EC-PCS always achieves the best area coverage.
Compared with GATE, RPSS performs worse in TEMPLE.
The reason is that the objective in TEMPLE has a much larger target area and a more complex structure than the objective in GATE.
Due to the fact that CPSS selects one photo from each cluster, it can guarantee a coarse-grained coverage.
Our photo selection scheme achieves a fine-grained coverage, thus performs better than the other methods, e.g., with a significant increase of $10.1\%$ and $5.6\%$ over RPSS and CPSS in the coverage ratio for the TEMPLE case, respectively.

\begin{footnotesize}
\begin{table}[!t]
\caption{3D reconstruction cost for the dataset of GATE.}
\label{tab:GATE_result}
\centering
\renewcommand\arraystretch{1.2}
\begin{tabular}{|m{0.65in}<{\centering}|m{0.44in}<{\centering}|m{0.44in}<{\centering}|m{0.53in}<{\centering}|m{0.5in}<{\centering}|}\hline
GATE                         &$\eta=80\%$            &$\eta=95\%$            &$\eta=99.5\%$         &$\eta=100\%$     \\\hline
\#Photos               &$22$                   &$38$                   &$51$                  &$68$             \\\hline
Monetary         &$0.203$ \$             &$0.654$ \$             &$1.361$ \$            &$1.981$ \$       \\\hline
\end{tabular}
\end{table}
\end{footnotesize}

\subsection{Performance for the Network Resource Allocation Scheme}
We evaluate our network resource allocation scheme for photo uploading by comparing it with 3 resource allocation schemes and 2 photo uploading schemes.
The experiments is based on the dataset TEMPLE.
\begin{itemize}
\item \emph{Fair Resource Allocation Scheme  (FRAS):} \\
The available bandwidth $B$ is equally shared by the $M$ edge participants.
\item \emph{Weighted Resource Allocation Scheme (WRAS):} \\
MEC server allocates the bandwidth according to each participant's weight, and the weight is set to be the number of photos for uploading.
\item \emph{Random Resource Allocation Scheme (RRAS):} \\
Allocate the bandwidth randomly to the edge participants.
\item \emph{Partial Photo Uploading to Cloud Server Scheme (PPU-CS):} \\
Edge participants upload part of raw photos to a remote cloud server based on a photo selection scheme.
The scheme is widely used in the cloud-based 3D reconstruction works, such as~\cite{SfM}.
\item \emph{Total Photo Uploading to Edge Server Scheme(TPU-ES):} \\
Edge participants upload all the raw photos to an edge server, and the photo selection is performed at the edge server.
The scheme is widely used in the edge-based crowdsensing works.
\end{itemize}

\begin{footnotesize}
\begin{table}[!t]
\caption{3D reconstruction cost for the dataset of TEMPLE.}
\label{tab:TEMPLE_result}
\centering
\renewcommand\arraystretch{1.2}
\begin{tabular}{|m{0.65in}<{\centering}|m{0.44in}<{\centering}|m{0.44in}<{\centering}|m{0.53in}<{\centering}|m{0.5in}<{\centering}|}\hline
TEMPLE                       &$\eta=80\%$            &$\eta=95\%$            &$\eta=99.5\%$         &$\eta=100\%$     \\\hline
\#Photos    &$37$                   &$64$                   &$99$                  &$141$             \\\hline
Monetary          &$0.543$ \$             &$1.434$ \$             &$3.137$ \$            &$6.577$ \$       \\\hline
\end{tabular}
\end{table}
\end{footnotesize}

\begin{figure}
  \begin{minipage}[t]{0.5\linewidth}
  \captionsetup{width=0.8\textwidth}
    \centering
    \includegraphics[width=1.6in]{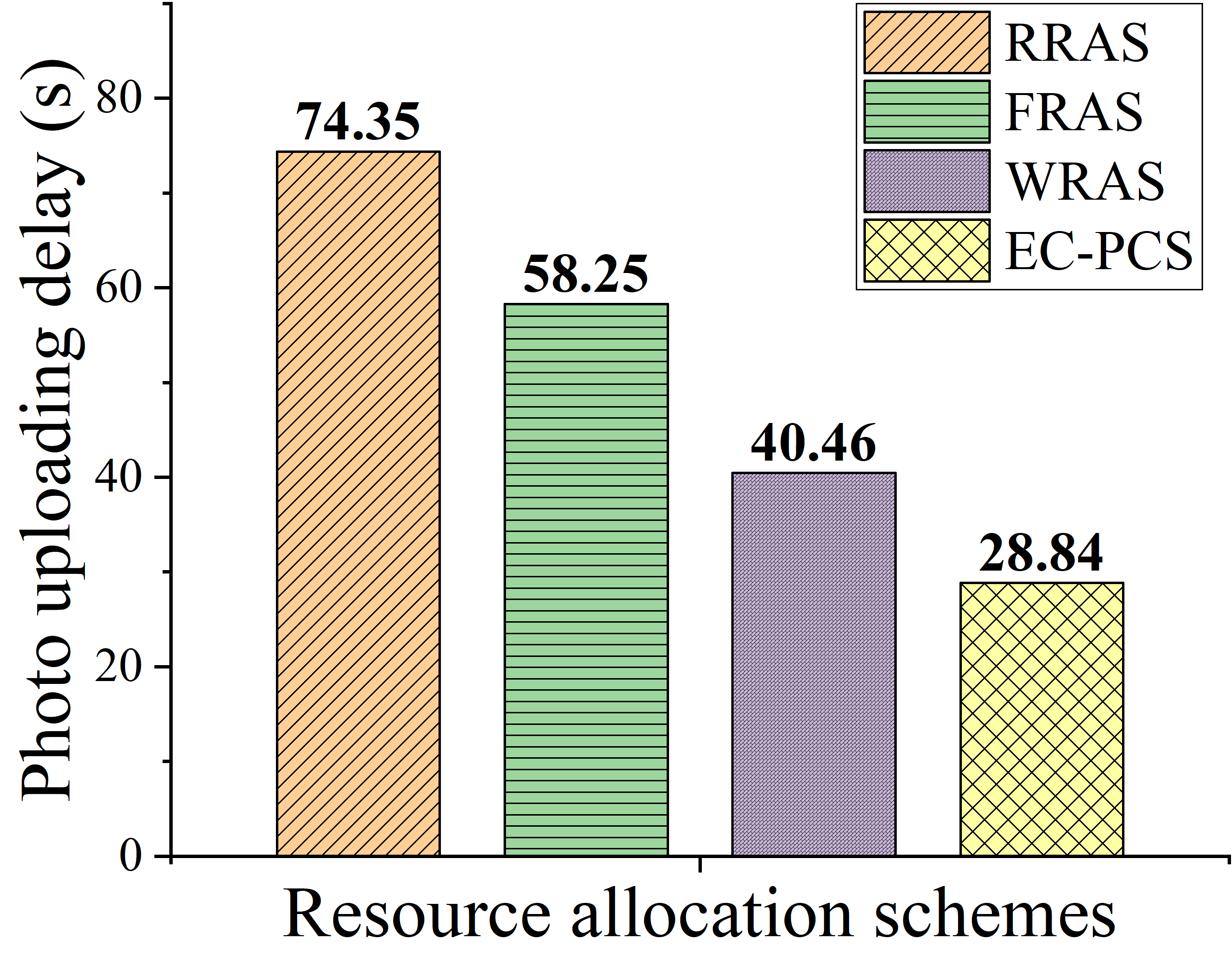}
    \caption{Photo uploading delay for different resource allocation schemes.}
    \label{fig:delay-RA}
  \end{minipage}%
  \begin{minipage}[t]{0.5\linewidth}
\captionsetup{width=0.8\textwidth}
    \centering
    \includegraphics[width=1.6in]{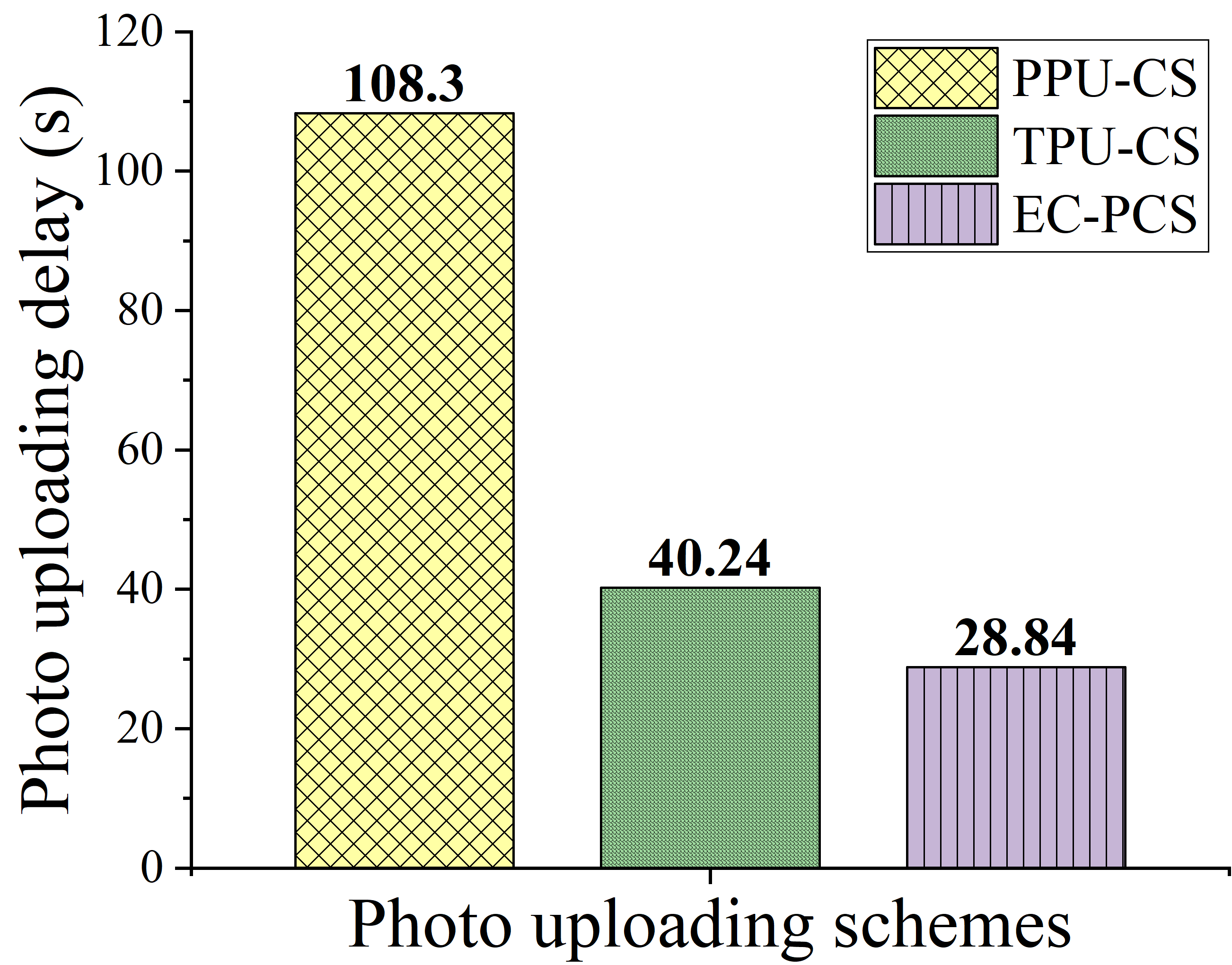}
    \caption{Photo uploading delay for different photo uploading schemes.}
    \label{fig:delay-PU}
  \end{minipage}
\end{figure}

Fig.~\ref{fig:delay-RA} reports the photo uploading delay for different resource allocation schemes.
Our network resource allocation scheme for EC-PCS obtains the best performance, because our scheme achieves the optimal solution as shown in Theorem 3 above.
Specifically, it can achieve more than $61\%$, $50.1\%$, and $29\%$ photo upload delay reduction over RRAS, FRAS and WRAS, respectively.
Fig.~\ref{fig:delay-PU} shows the photo uploading delay for different photo uploading schemes.
Note that traditional cloud-based photo uploading schemes performs much worse than the edge-based schemes.
The reason is that cloud-based schemes not only upload photos to the edge, but also transmit the photos to remote cloud server.
On the other hand, compared with the total uploading scheme TPU-ES, we observe a pre-selection scheme (i.e., partial uploading) is useful to reduce the uploading delay, achieving $28\%$ delay reduction over the TPU-ES scheme.
Indeed, the photo uploading scheme for our EC-PCS is an edge-based partial uploading method, thus can minimize the photo uploading delay.

At last, we qualitatively show the effectiveness of our EC-PCS by comparing it with three existing real-time 3D reconstruction approaches, 3D modeling on the go~\cite{7335496}, Photo Tourism~\cite{SfM} and Photo City~\cite{PhotoCity}, respectively.
In this paper, photo collection delay refers to the sum of photo selection delay and photo uploading delay.
As reported in TABLE.~\ref{tab:collection_delay}, our EC-PCS can reduce the photo collection delay greatly for the following reasons: i) Collecting photos directly from ubiquitous mobile and Internet of Things devices, ii) Uploading photos through high-speed MEC network, and iii) Processing photos at edge server, instead of transmitting to the remote cloud server.

\begin{footnotesize}
\begin{table}[!t]
\caption{Photo collection delay comparisons.}
\label{tab:collection_delay}
\centering
\renewcommand\arraystretch{1.2}
\begin{tabular}{|m{1.5in}<{\centering}|m{1.5in}<{\centering}|}\hline
3D Reconstruction Schemes            &Photo Collection Delay            \\\hline
EC-PCS                               &a few seconds to tens of second   \\\hline
3D modeling on the go~\cite{7335496} &ten minutes                       \\\hline
Photo Tourism~\cite{SfM}             &a few days                       \\\hline
Photo City~\cite{PhotoCity}          &a few weeks                       \\\hline
\end{tabular}
\end{table}
\end{footnotesize}

\section{Conclusion}{\label{sec:Conclusion}}
In this paper, we address the issue of real-time 3D reconstruction in 5G multi-access edge computing (MEC) environments.
To this end, we propose a novel edge computing-based photo crowdsourcing (EC-PCS) framework and show the main modules of the framework.
The core functions of the framework are i) select a proper set of photos that best cover the target area with minimum monetary cost, and ii) allocate the limited network resources to the participants, in order to minimize the uploading delay for the selected photos.
In view of this, we first present a photo selection scheme, and prove that the problem of finding the optimal set of photos is NP-hard.
Then, in order to minimize the uploading delay for the selected photos, we propose a network resource allocation scheme, and obtain the optimal resource allocation strategy through KKT conditions.
Last but not least, we introduce the 3D reconstruction module and the 3D model caching module of our framework.
Both simulation and experimental results are based on real-world datasets to demonstrate the performance of our EC-PCS framework.
In the future, we will address a number of interesting open questions and directions to extend our EC-PCS, such as i) classify participants into several categories (according to their mobility or capacity), ii) cost-effective photo crowdsourcing iii) auction-based photo selection and iv) privacy-preserving photo transmission (e.g., differential privacy and homomorphic encryption based photo aggregation).

\section*{Proof A: proof of the Theorem~\ref{prop:nphard}}{\label{app:appA}}

\begin{proof}
Given a collection $\mathcal{P}^{*}$ of subsets, it is obviously that we can check the if the union of them can cover at least $\eta$ (in \%) elements of $\mathcal{A}^{Tar}$ in polynomial time.
Thus, the cost-aware photo selection problem is in NP.
Then, we prove that finding the optimal solution is NP-hard.
To prove that the cost-aware photo selection problem is NP Hard, we consider a problem which has already been proven to be NP-Hard, and prove that this problem can be reduced to our cost-aware photo selection problem.
To this end, we consider the vertex cover problem (VCP), which is NP-complete (and hence NP-hard), and prove VCP $\leq_{p}$ cost-aware photo selection problem.

Vertex Cover Problem (VCP): Given a graph $\mathcal{G}=(\mathcal{V},\mathcal{E})$ and a positive integer $K$, if there is a subset $V$ of vertices of size at most $K$, such that every edge in the graph is connected to some vertex in $\mathcal{V}$.

We first define function $f$ as follows:

\begin{equation}{\label{eq:reduction}}
\begin{aligned}
<\mathcal{E},\mathcal{C},K>=f(\mathcal{G},K),\\
\mathcal{C}=\{C_{v}, v\in \mathcal{V}\},\\
C_{v}=\{e\in\mathcal{E},v\in e\}.
\end{aligned}
\end{equation}

Assume $\mathcal{G}$ has a vertex cover $W$ of size $K$, and let $\mathcal{D}=\{C_{v}, v\in W$\}, thus $|\mathcal{D}|=K$.
Note that $\mathcal{D}$ is a $\mathcal{C}$-cover of $\mathcal{E}$.
To justify, we assume $e={u,v}\in \mathcal{E}$, then, $e$ is in both $C_{u}$ and $C_{v}$.
However, since $W$ is a vertex cover, at least one of $u$ or $v$ is in $W$.
Therefore, at least one of the sets $C_{u}$ or $C_{v}$ is in $\mathcal{D}$.

Conversely, assume $\mathcal{E}$ has a $\mathcal{C}$-cover $D$ of size $K$, let $W=\{v\in\mathcal{V},C_{v}\in\mathcal{D}\}$.
Then, $|W|=K$.
Note that $W$ is a vertex cover of of $\mathcal{G}$.
To justify, we assume $e\in\mathcal{E}$.
Since $\mathcal{D}$ is a $\mathcal{C}$-cover of $\mathcal{E}$.
There is a set $C_{v}\in \mathcal{D}$ such that $e\in C_{v}$
The definition of $W$ then says that $v\in W$, and the definition of $C_{v}$ says that $v\in e$.
Thus, the function $f$ is a polynomial-time reduction of VCP to the cost-aware photo selection problem.

\end{proof}

\section*{Proof B: proof of the Theorem~\ref{prop:approximation}}{\label{app:appC}}

Let $A_{i}$ denote the number of grids uncovered at iteration $i$.
Thus there are $|\mathcal{A}^{Tar}|-A_{i}$ covered grids at iteration $i$ and another $A_{i}^{'}=A_{i}-(|\mathcal{A}^{Tar}|-\lceil \eta\cdot|\mathcal{A}^{Tar}| \rceil)$ grids to cover.
We will take the set corresponding to the grid $g_{j}$ that minimizes $\frac{G_{m,n}}{\min\{|\mathcal{A}^{Tar}_{m,n}\cap A_{i}|,A_{i}^{'}\}}$
Let $c_{i}$ be the cost charged to $g_{i}$, so if $g_{i}$ belongs to $\mathcal{A}^{Tar}_{m,n}$ we have $c_{i}=\frac{G_{m,n}}{\min\{|\mathcal{A}^{Tar}_{m,n}\cap A_{i}|,A_{i}^{'}\}}$.
We have:
\begin{equation}
\begin{aligned}
&c_{i}=\frac{G_{m,n}}{\min\{|\mathcal{A}^{Tar}_{m,n}\cap A_{i}|,A_{i}^{'}\}}\leq \frac{G_{j,k}}{\min\{|\mathcal{A}^{Tar}_{j,k}\cap A_{i}|,A_{i}^{'}\}},\\
&|\mathcal{A}^{Tar}_{j,k}\cap A_{i}|>0\\
\end{aligned}
\end{equation}
Thus,
\begin{equation}
\begin{aligned}
&\sum_{j\in\mathcal{P}^{*}}\geq \frac{G_{m,n}}{\min\{|\mathcal{A}^{Tar}_{m,n}\cap A_{i}|,A_{i}^{'}\}}\cdot \sum_{j\in\mathcal{P}^{*}}\min\{|\mathcal{A}^{Tar}_{j,k}\cap A_{i}|,A_{i}^{'}\}\\
&=c_{i}\cdot \sum_{j\in\mathcal{P}^{*}}\min\{|\mathcal{A}^{Tar}_{j,k}\cap A_{i}|,A_{i}^{'}\}\leq c_{i}\cdot A_{i}^{'}.\\
\end{aligned}
\end{equation}

Note that $\cup_{p_{m,n}\in \mathcal{P}^{*}}\mathcal{A}^{Tar}_{m,n}\cap A_{i}$ covers at least $A_{i}^{'}$ grids.
Therefore, $c_{i}\leq \frac{V^{*}}{A_{i}^{'}}\leq \frac{V^{*}}{i}$.
To sum up, the total cost of Algorithm~\ref{alg:greedy} can be summarized as follows:
\begin{equation}
\begin{aligned}
\text{cost}=&c_{\lceil \eta\cdot|\mathcal{A}^{Tar}| \rceil}+...+c_{1}=V^{*}\cdot(\sum_{i=1}^{\lceil \eta\cdot|\mathcal{A}^{Tar}| \rceil}\frac{1}{i}),\\
&=F_{\lceil \eta\cdot|\mathcal{A}^{Tar}| \rceil}\cdot V^{*}.\\
\end{aligned}
\end{equation}
Thus, Algorithm~\ref{alg:greedy} for the cost-aware photo selection problem achieves an approximation of $F_{\lceil \eta\cdot|\mathcal{A}^{Tar}| \rceil}$.

\section*{Proof C: proof of the Theorem~\ref{prop:convex}}{\label{app:appB}}
We define the Lagrangian for the problem~(\ref{eq:uploading_delay_convert}) as follows:
\begin{equation}
\begin{aligned}
\label{eq:Lagrangian}
L(\mathcal{B},t,\boldsymbol{u},v)=t+\sum_{i=1}^{M}u_{i}h_{i}(\mathcal{B},t)+vg(\mathcal{B}).
\end{aligned}
\end{equation}
The Karush-Kuhn-Tucker (KKT) conditions~\cite{Ye2014} can be written as:
\begin{equation}
\begin{aligned}
\label{eq:KKT}
&1-\sum_{i=1}^Mu_{i}=0,\quad\sum_{i=1}^{M}u_{i}\nabla_{\mathcal{B}} h_{i}(\mathcal{B},t)+v\nabla_{\mathcal{B}}g(\mathcal{B})=\mathbf{0},\\
&u_{i}\cdot h_{i}(\mathcal{B},t)=0, \quad i=1,2,...,M,\\
&h_{i}(\mathcal{B},t)\leq 0, g(\mathcal{B})=0, \quad i=1,2,...,M,\\
&u_{i}\geq 0, \quad i=1,2,...,M,
\end{aligned}
\end{equation}
where
\begin{align}
\nabla_{\mathcal{B}} h_{i}(\mathcal{B},t)&=\left[0,\cdots,-\frac{\sum_{P_{i,n}\in \mathcal{P}^{*}} D_{i,n}}{B^2_{i}\log_{2}\left(1+\mathbb{SNR}_{i}\right)},\cdots,0\right]^T,
\nonumber\\
\nabla_{\mathcal{B}}g(\mathcal{B})&=\mathbf{1}.
\nonumber
\end{align}
Putting $\nabla_{\mathcal{B}} h_{i}(\mathcal{B},t)$ and $\nabla_{\mathcal{B}}g(\mathcal{B})$ into \eqref{eq:KKT}, we can obtain the optimal resource allocation policy $\mathcal{B}^{*}$ as follows:
\begin{equation}
\begin{aligned}
\mathcal{B}^{*}&=\{B_{1}^{*},B_{2}^{*},...,B_{M}^{*}\},\\
B_{m}^{*}&=\frac{B\cdot\frac{\sum_{P_{m,n}\in \mathcal{P}^{*}} D_{m,n}}{\log_{2}(1+\mathbb{SNR}_{m})}}{\sum_{m=1}^{M}\left(\frac{\sum_{P_{m,n}\in \mathcal{P}^{*}} D_{m,n}}{\log_{2}(1+\mathbb{SNR}_{m})}\right)},\quad \forall m.
\end{aligned}
\end{equation}
The proof is thus completed.


%



\ifCLASSOPTIONcaptionsoff
  \newpage
\fi



%

\section*{Acknowledgment}
This work was supported in part by the National Science Foundation of China (No. U1711265, No. 61972432, No. U1911201); the Program for Guangdong Introducing Innovative and Entrepreneurial Teams (No.2017ZT07X355);the Pearl River Talent Recruitment Program (No.2017GC010465); Guangdong Special Support Program (No. 2017TX04X148); the Fundamental Research Funds for the Central Universities (No. 20lgpy135); Guangdong Basic and Applied Basic Research Foundation (No. 2019A1515010030).

\bibliographystyle{IEEEtran}

\bibliography{bibThesis}

\end{document}